\documentclass[a4paper,preprint]{revtex4}

\begin{document}
\title{Exact Result for the Nonlocal Conserved Kardar-Parisi-Zhang Equation}
\author{Eytan Katzav}
\email{eytak@post.tau.ac.il} \affiliation {School of Physics and
Astronomy, Raymond and Beverly Sackler Faculty of Exact Sciences,
Tel Aviv University, Tel Aviv 69978, Israel}

\begin{abstract}
I analyze the Nonlocal Conserved Kardar-Parisi-Zhang (NCKPZ)
equation with spatially correlated noise. This equation is also
known as the Nonlocal Molecular Beam Epitaxy (NMBE) equation andv
was originally suggested to study the effect of the long-range
nature of interactions coupled with spatially correlated noise on
the dynamics of a volume conserving surface. I find an exact
result for a subfamily of NCKPZ models in one dimension, and thus
establish an exact result for MBE processes for the first time.
Then, to complete the picture, I construct a Self-Consistent
Expansion (SCE) and get results that are consistent with the exact
result in one dimension. I conclude by discussing previous results
obtained for NCKPZ using dynamic renormalization group approach,
and find that this approach does not yield the exact result I
obtain. This discussion shows the advantage of the using SCE to
deal with non-linear stochastic equations.
\end{abstract}

\maketitle

The field of disorderly surface growth has received much attention
during the last two decades. A multitude of different phenomena
such as fluid flow in porous media, propagation of flame fronts,
flux lines in superconductors not to mention deposition processes,
bacterial growth and "DNA walk" \cite{barabasi95} are all said to
be related to the famous Kardar-Parisi-Zhang (KPZ) equation
\cite{kpz86}.

In spite of the great success of KPZ to describe many growth
models and phenomena, there has been a growing pool of data that
is not well described by KPZ, that called for further research.
One of the first classes that belongs to this non KPZ behavior is
the well known conserved KPZ equation class (sometimes also
called the Molecular-Beam-Epitaxy (MBE) class)
\cite{wolf90}-\cite{sun93}. However, the modified behavior
introduced by the MBE equation is not at all sufficient to account
for all the rich non KPZ experimental data in the field.

A different line of research suggested that the basic growth
equation should not be changed. Instead, the white noise that
appears in the original KPZ equation should be either temporally
or spatially correlated \cite{medina89}-\cite{katzav99}. This
approach was indeed quite successful, but it still failed to give
a good account for all the measured scaling exponents.

Recently, Mukherji and Bhattacharjee \cite {Mukh97} proposed a
phenomenological equation that takes into account the presence of
long-range interactions. This equation, known as the Nonlocal KPZ
(NKPZ) equation, has a nonlinear term that couples the gradients
of the growing surface at two different points. This equation was
studied using the Dynamic Renormalization Group (DRG) method \cite
{Mukh97} and so was its generalization that includes spatially
correlated noise \cite{chat99}. It turned out that a very rich
family of phases was found.

Just like in the local case, a nonlocal conserved KPZ (NCKPZ)
equation was needed in order to account for the dynamics of a
volume conserving surface that experiences long-range
interactions. This equation was proposed by Jung {\it{et al}}.
\cite{Jung98, Jung2000}.  The continuum equation they suggest for
the height of the surface $h\left( {\vec r,t} \right)$ at a point
$\vec r$ and time $t$ measured relative to its spatial average is
\begin{eqnarray}
 \frac{{\partial h\left( {\vec r,t} \right)}}{{\partial t}} &=&
   - K\nabla ^4 h\left( {\vec r,t} \right) - \frac{1}{2}\nabla ^2 \int {d^dr'g\left( {\vec r'} \right)}
  \nonumber\\
   && \times \nabla h\left( {\vec r + \vec r',t} \right) \cdot \nabla h\left( {\vec r - \vec r',t} \right) + \eta \left( {\vec r,t} \right)
 \label{1},
\end{eqnarray}
where $K$ is a constant, the kernel $g\left( {\vec r} \right)$
represents the long-range interactions with a short-range part
$\lambda _0 \delta ^d \left( {\vec r} \right)$ and a long-range
part $ \sim \lambda _\rho  r^{\rho  - d} $, where d is the
substrate dimension and $\rho$ is an exponent characterizing the
decay of the long range interaction. More precisely in Fourier
space $\hat g\left( q \right) = \lambda _0  + \lambda _\rho  q^{ -
\rho } $. Finally, $\eta \left( {\vec r,t} \right)$ is a
conservative spatially correlated noise term that satisfies
\begin{eqnarray}
 \left\langle {\eta \left( {\vec r,t} \right)} \right\rangle  &=& 0
\nonumber\\
 \left\langle {\eta \left( {\vec r,t} \right)\eta \left( {\vec r',t} \right)} \right\rangle  &\sim& \nabla^2 \left| {\vec r - \vec r'} \right|^{2\sigma  - d} \delta \left( {t - t'} \right)
\label{2},
\end{eqnarray}
where $\sigma$ is an exponent characterizing the decay of spatial
correlations (in the case of conservative white noise , which is
of great interest in the MBE system, $\sigma = 0$).

After suggesting the model, Jung {\it{et al}}.
\cite{Jung98,Jung2000} investigated the NCKPZ problem using DRG
analysis, and derived a very rich and complex picture of phases -
each one of them described by a different set of critical
exponents. In this letter I will not go into the details of all
those phases, but would like to focus on the strong coupling
behavior of NCKPZ. In this regime the DRG analysis yields the
power counting result
\begin{equation}
\alpha  = {{\left( {2 - d - \rho  + 2\sigma } \right)}
\mathord{\left/
 {\vphantom {{\left( {2 - d - \rho  + 2\sigma } \right)} 3}} \right.
 \kern-\nulldelimiterspace} 3},\quad \quad z = {{\left( {d + 10 - 2\rho  - 2\sigma } \right)} \mathord{\left/
 {\vphantom {{\left( {d + 10 - 2\rho  - 2\sigma } \right)} 3}} \right.
 \kern-\nulldelimiterspace} 3}
 \label{DRG},
\end{equation}
where $\alpha$ is the roughness exponent and $z$ is the dynamic
exponent. This solution is said to valid for positive values of
$\rho$ and for $d<2+2\rho+2\sigma$.

In this letter I will show that the above DRG analysis does not
give the full description of the different phases in the strong
coupling regime of NCKPZ. More specifically, I will show that an
exact solution is possible for some one-dimensional cases (namely
for $\rho=2+2\sigma$). It is interesting to mention that this is
the first time an exact result is suggested for the family of MBE
processes. It will then be immediately seen that this exact
solution is not compatible with the DRG result. Then, in order to
clarify the situation, I will apply a method developed by Schwartz
and Edwards \cite{SE92,SE98,katzav99} (also known as the
Self-Consistent-Expansion (SCE) approach). This method has been
previously applied successfully to the KPZ equation. The method
gained much credit by being able to give a sensible prediction for
the KPZ critical exponents in the strong coupling phase, where, as
is well known, many Renormalization-Group (RG) approaches failed
(as well as DRG of course) \cite{Natterman}. For the specific
problem of NCKPZ, the SCE method indicates that two strong
coupling solutions exist, one of which agrees with the power
counting solution also obtained by DRG, while the other one
recovers the exact one-dimensional result (and also extends it to
higher dimensions). It turns out that each solution is valid for
different values of the parameters $\sigma$, $\rho$ and $d$
(mutually excluding values).

In order to derive the exact result, I have to consider eqs.
(\ref{1})-(\ref{2}) in Fourier components

\begin{equation}
\frac{{\partial h_q }}{{\partial t}} =  - Kq^4 h_q  -
\frac{1}{{\sqrt \Omega  }}\sum\limits_{\ell ,m} {\left( {\lambda
_0  + \lambda _\rho  q^{ - \rho } } \right)q^2 \left( {\vec \ell
\cdot \vec m} \right)\delta _{q,\ell  + m} h_\ell  h_m }  + \eta
_q
\label{3},
\end{equation}
where $\eta _q$ is the Fourier component of the noise that
satisfies:
\begin{eqnarray}
 \left\langle {\eta _q \left( t \right)} \right\rangle  &=& 0 \nonumber\\
 \left\langle {\eta _q \left( t \right)\eta _{q'} \left( {t'} \right)} \right\rangle  &=& 2D_0 q^{ 2- 2\sigma } \delta_{q+q'} \delta \left( {t - t'} \right)
\label{4}.
\end{eqnarray}

I will show that for the case $\lambda _0  = 0$ an exact solution
is possible for the steady state distribution resulting from eq.
(\ref{3}). However, before doing that I would like to mention that
when $\rho > 0$, the $\lambda _\rho $ term in eq. (\ref{3})
dominates over the $\lambda _0 $ term in the asymptotic small q
limit, so that this term can be neglected. On the other hand when
$\rho < 0$ the $\lambda _\rho $ term can be neglected compared to
the $\lambda _0 $ term, and I retain the standard MBE results,
unless $\lambda _0 = 0$ exactly (The specific case $\lambda _0  =
0$ and $\rho < 0$ is reasonable only under special physical
conditions \cite{Villain}). It will be understood therefore, that
while the results obtained for positive $\rho$'s hold also for
$\lambda _0 \neq 0$, the results obtained for negative $\rho$'s
hold only when $\lambda _0$ is strictly zero. In other cases (i.e.
$\rho<0$ and $\lambda _0 \neq 0$) the MBE results are recovered.

I now turn from the NCKPZ equation in Langevin form to a
Fokker-Planck form

\begin{eqnarray}
 \frac{{\partial P}}{{\partial t}} &+& \sum\limits_q {\frac{\partial }{{\partial h_q }}\left[ {D_0 q^{ 2- 2\sigma } \frac{\partial }{{\partial h_{ - q} }} + Kq^4 h_q  + } \right.}  \nonumber\\
 &&\left. {\quad \quad \quad \quad  + \frac{{\lambda _\rho  q^{2 - \rho } }}{{\sqrt \Omega  }}\sum\limits_{\ell ,m} {\left( {\vec \ell  \cdot \vec m} \right)\delta _{q,\ell  + m} h_\ell  h_m } } \right]P = 0
\label{5},
\end{eqnarray}
(where $P\left( \{h_q\} ,t \right)$ is the probability of having
the configuration $\left\{ {h_q} \right\} $ at a specific time t).

Consider now the family of models defined by $\rho=2+2\sigma$. By
direct substitution, it can be verified immediately that the
following probability function
\begin{equation}
P_0  = N\exp \left[ { - {\textstyle{1 \over 2}}\sum\limits_q
{\frac{{h_q h_{ - q} }}{{\phi _q }}} } \right]
\label{6},
\end{equation}
(where N is the normalization yielding $\int {P_0 \left\{ h
\right\}Dh = 1}$, and $\phi _q $ is the two point correlation
function) is a solution of the steady-state Fokker-Plack equation
(i.e. with ${{\partial P} \mathord{\left/ {\vphantom {{\partial P}
{\partial t}}} \right.
 \kern-\nulldelimiterspace} {\partial t}} = 0$) for the one-dimensional
 case $\left( {d = 1} \right)$ if $\phi _q  = Aq^{ - \Gamma } $
 with $A = {{D_0 } \mathord{\left/
 {\vphantom {{D_0 } K}} \right. \kern-\nulldelimiterspace} K}$ and
 $\Gamma  = \rho $. Translating this result to the frequently used notation
 in the field of surface growth I get

 \begin{equation}
\alpha  = \frac{\rho-1}{2}
 \label{7},
 \end{equation}
 where $\alpha$ is just the roughness exponent. This result
 is therefore consistent with the known exact solution of one-dimensional KPZ
 $\alpha  = {1 \mathord{\left/ {\vphantom {1 2}} \right.
 \kern-\nulldelimiterspace} 2}$ (by setting $\sigma  =  0$ and $\rho  = 2$).
 However, this result is manifestly different from the DRG result
 cited in eq. (\ref {DRG}).

After establishing the exact result I turn to the application of
the SCE method to NCKPZ in order to clarify the conflict as well
as to complete the picture for any noise exponent $\sigma $ and
for higher dimensions. The SCE method is based on the
Fokker-Planck form of NCKPZ equation, which is the reasonable
thing to do once you have an exact solution for that equation in
one dimension. The method constructs a self-consistent expansion
of the distribution of the field concerned. The expansion is
formulated in terms of $\phi _q $ and $\omega_q$, where $\phi _q $
is the two-point function in momentum space, defined by $\phi _q =
\left\langle {h_q h_{ - q} } \right\rangle _S $, (the subscript S
denotes steady state averaging), and $\omega _q $ is the
characteristic frequency associated with $h_q $. I expect that for
small enough q, $\phi _q $ and $\omega _q $ obey power laws in q,

\begin{equation}
\phi _q  = Aq^{ - \Gamma } \quad \quad and\quad \quad \omega _q  =
Bq^\mu \label{8}.
\end{equation}

[Since dynamic surface growth is a remarkably multidisciplinary
field, there are almost as many notations as there are workers in
the field. Therefore I give a brief translation of our notations
to those most frequently used:
\begin{equation}
 z = \mu,\quad \alpha  = {{\left( {\Gamma  - d} \right)}
\mathord{\left/
 {\vphantom {{\left( {\Gamma  - d} \right)} 2}} \right.
 \kern-\nulldelimiterspace} 2}\quad \quad and\quad \quad \beta  = {{\left( {\Gamma  - d} \right)} \mathord{\left/
 {\vphantom {{\left( {\Gamma  - d} \right)} {2\mu }}} \right.
 \kern-\nulldelimiterspace} {2\mu }}
\label{9}.]
\end{equation}

The SCE method produces, to second order in this expansion, two
nonlinear coupled integral equations in $\phi _q $ and $\omega _q
$, that can be solved exactly in the asymptotic small q limit to
yield the required scaling exponents governing the steady state
behavior and the time evolution.

By defining $M_{q\ell m}  = \frac{{\lambda _\rho  }}{{\sqrt \Omega
}}q^{2 - \rho } \left( {\vec \ell  \cdot \vec m} \right)\delta
_{q,\ell  + m} $, $K_q  = Kq^4 $ and $D_q  = D_0 q^{ 2- 2\sigma }
$ it can be seen that the NCKPZ equation is of the general form
discussed in refs. \cite{SE92,SE98}, where the SCE method is
derived. Working to second order in the expansion, this yields the
two coupled non-linear integral equations

\begin{eqnarray}
 D_q  &-& K_q \phi _q  + 2\sum\limits_{\ell ,m} {\frac{{M_{q\ell m} M_{q\ell m} \phi _\ell  \phi _m }}{{\omega _q  + \omega _\ell   + \omega _m }}}  -
 \nonumber\\
   &-& 2\sum\limits_{\ell ,m} {\frac{{M_{q\ell m} M_{\ell mq} \phi _m \phi _q }}{{\omega _q  + \omega _\ell   + \omega _m }}}  - 2\sum\limits_{\ell ,m} {\frac{{M_{q\ell m} M_{m\ell q} \phi _\ell  \phi _q }}{{\omega _q  + \omega _\ell   + \omega _m }}}  = 0
\label{10},
\end{eqnarray}

and
\begin{equation}
K_q  - \omega _q  - 2\sum\limits_{\ell ,m} {M_{q\ell m}
\frac{{M_{\ell mq} \phi _m  + M_{m\ell q} \phi _\ell  }}{{\omega
_\ell   + \omega _m }}}  = 0
\label{11},
\end{equation}
where in deriving the last equation I have used the Herring
consistency equation \cite{Herring}. In fact Herring's definition
of $\omega _q $ is one of many possibilities, each leading to a
different consistency equation. But it can be shown, as previously
done in ref. \cite{SE98}, that this does not affect the exponents
(universality).

A full solution of equations (\ref{10}) and (\ref{11}) in the
limit of small q's (i.e. large scales) yields a very rich family
of solutions that will not interest us here, but may be the
subject of a future detailed paper. Instead, I focus on the strong
coupling solutions obtained by SCE. As mentioned above, two strong
coupling solutions are obtained by SCE, each of them describes the
critical exponents for a different set of the parameters $\rho$,
$\sigma$ and $d$.

The first solution is obtained just by power counting, and reads
$\mu = {{\left( {d + 10 - 2\rho  - 2\sigma } \right)}
\mathord{\left/ {\vphantom {{\left( {d + 10 - 2\rho  - 2\sigma }
\right)} 3}} \right. \kern-\nulldelimiterspace} 3}$ and $\Gamma  =
{{\left( {d + 4 - 2\rho  + 4\sigma } \right)} \mathord{\left/
{\vphantom {{\left( {d + 4 - 2\rho  + 4\sigma } \right)} 3}}
\right. \kern-\nulldelimiterspace} 3}$. This solution is valid for
$\rho - 1 < \sigma $, $d < 6 - 2\sigma  + \rho $, $d < 2 + 2\sigma
+ 2\rho $ and $d < 8 + 2\sigma  - 4\rho $. Therefore, by direct
inspection, this solution is not valid for the one-dimensional
case with $\rho=2+2\sigma$, and hence does not recover the exact
Gaussian result obtained above. It should also be mentioned that
when translated to frequently used notation (using eq. (\ref{9}))
it can be seen that this solution is actually the same as the one
obtained by the DRG analysis as in eq. (\ref{DRG}).

The second solution is a non power counting solution. It turns out
that for $d = 1$ eq. (\ref {10}) is exactly solvable, and yields
$\Gamma  = \rho $. This result is actually the exact Gaussian
solution of the one-dimensional case, mentioned above. In
addition, from eq. (\ref {11}), the dynamical exponent can be
extracted for a certain range of the parameters, namely $1 < \rho
< 3$ and $\rho  > {\left( {4\sigma + 5} \right)}$, where it
becomes $\mu = {\left( {9 - 3\rho } \right)}$.

For dimensionality higher than one (i.e. $d\geq2$) such an exact
solution in closed form cannot be found, and the strong coupling
solution is determined from the combination of the scaling
relation $\mu  = {{\left( {d + 8 - \Gamma  - 2\rho } \right)}
\mathord{\left/ {\vphantom {{\left( {d + 8 - \Gamma  - 2\rho }
\right)} 2}} \right. \kern-\nulldelimiterspace} 2}$, and the
transcendental equation $F\left( {\Gamma ,\mu } \right) = 0$,
where F is given by
\begin{eqnarray}
 F\left( {\Gamma ,\mu } \right) =  &-& \int {d^d t\frac{{\vec t \cdot \left( {\hat e - \vec t} \right)}}{{t^\mu   + \left| {\hat e - \vec t} \right|^\mu   + 1}}\left[ {\left( {\hat e \cdot \vec t} \right)\left| {\hat e - \vec t} \right|^{2 - \rho } t^{ - \Gamma }  + \hat e \cdot \left( {\hat e - \vec t} \right)t^{2 - \rho } \left| {\hat e - \vec t} \right|^{ - \Gamma } } \right]}  + \nonumber\\
 &+& \int {d^d t\frac{{\left[ {\vec t \cdot \left( {\hat e - \vec t} \right)} \right]^2 }}{{t^\mu   + \left| {\hat e - \vec t} \right|^\mu   + 1}} t^{ - \Gamma } \left| {\hat e - \vec t} \right|^{ - \Gamma }}
\label{12},
\end{eqnarray}
and $\hat e$ is a unit vector in an arbitrary direction. The last
equation has to be solved numerically of course. It addition, This
solution is valid as long as the solutions of the last equations
satisfy the following four conditions: $d + 4 - 3\Gamma - 2\rho  +
4\sigma  < 0$, $d - 3\Gamma  + 2\rho  < 0$, $d < \Gamma $ and $d -
4 - \Gamma  + 2\rho  < 0$.

It should be mentioned that SCE yields many other "weak coupling"
solutions to NCKPZ. They are called "weak coupling" in the sense
that all of them yield the exponents of the linear MBE problem
with correlated noise in the limit of $\rho=0$. These solutions
are also not obtained by DRG.

In summary, I have established an exact solution for the
one-dimensional NCKPZ equation in the special case where $\rho  =
2 + 2\sigma $. Then, I calculated the critical exponents for the
most general case (i.e. any $\rho $, $\sigma $
 and $d$) using the SCE method, and obtained two possible strong coupling solutions.
 The first strong coupling solution is a power counting one, and is
 identical to the previous result of the DRG analysis.
 The second strong coupling solution is non power counting,
 and is not obtained by DRG. For $d=1$, this solution turns out to be the
 exact Gaussian solution obtained at the beginning. This suggest
 that there is a big advantage in using the SCE to deal with such
 nonlinear problems.

Another advantage of using SCE has to do with the well known
technical problem raised by Janssen \cite{Janssen97}. Janssen
suggested that there should be a correction to the scaling
relation because of a nontrivial renormalization of the coupling
constant. This correction appears in a two-loop RG calculation,
but not in a one-loop calculation. Janssen himself suspected that
in order to avoid this difficulty, a mode-coupling approach might
be more appropriate. The SCE method is indeed closely related to
the mode-coupling approaches, in the sense that similar (but not
identical) equations are obtained, while the underlying derivation
is different. Therefore, the fact SCE recovers the one-loop DRG
power counting results of Jung {\it {et al}}. , for a specific
range of the parameters \cite{Jung98,Jung2000}, is a nontrivial
corroboration of that result.

Acknowledgement: I would like to thank Moshe Schwartz for useful
discussions.

\end{document}